\documentclass[10pt,conference]{IEEEtran}
\IEEEoverridecommandlockouts

\makeatletter
\newcommand{\biggg}[1]{{\hbox{$\left#1\vbox to 20.5pt{}\right.\n@space$}}}
\newcommand{\Biggg}[1]{{\hbox{$\left#1\vbox to 23.5pt{}\right.\n@space$}}}
\newcommand{\bigggg}[1]{{\hbox{$\left#1\vbox to 26.5pt{}\right.\n@space$}}}
\newcommand{\Bigggg}[1]{{\hbox{$\left#1\vbox to 29.5pt{}\right.\n@space$}}}
\newcommand{\biggggg}[1]{{\hbox{$\left#1\vbox to 32.5pt{}\right.\n@space$}}}
\newcommand{\Biggggg}[1]{{\hbox{$\left#1\vbox to 35.5pt{}\right.\n@space$}}}
\newcommand{\bigggggg}[1]{{\hbox{$\left#1\vbox to 38.5pt{}\right.\n@space$}}}
\newcommand{\Bigggggg}[1]{{\hbox{$\left#1\vbox to 41.5pt{}\right.\n@space$}}}
\makeatother

\usepackage{makecell}
\usepackage{booktabs}
\usepackage{bm,cite,algorithm,algorithmic,float,amsmath,amssymb}
\usepackage[dvips]{graphicx}
\usepackage{subfigure,amsfonts}
\usepackage{mdwmath}
\usepackage{mdwtab}
\usepackage{xcolor}
\usepackage{cool}
\usepackage{makeidx,enumitem}
\usepackage{balance}
\floatname{algorithm}{Algorithm}

\makeatletter
\renewcommand\paragraph{\@startsection{paragraph}{4}{\z@}%
            {-2.5ex\@plus -1ex \@minus -.25ex}%
            {1.25ex \@plus .25ex}%
            {\normalfont\normalsize\itshape}}
\makeatother
\usepackage{epstopdf}

\begin{document}
\title{Band Assignment in Dual Band Systems: A Learning-based Approach \thanks{Part of this work was supported financially by the National Science Foundation.}
 \thanks{The authors are at the Ming Hsieh Department of Electrical Engineering, University of Southern California, 
Los Angeles, CA 90089, USA
(email: burghal,wang78,molisch@usc.edu).}}
\author{
\IEEEauthorblockN{Daoud Burghal, \emph{Student Member, IEEE,} Rui Wang, \emph{Student Member, IEEE,} and Andreas F. Molisch \emph{Fellow, IEEE}}}
\maketitle

\begin{abstract}
We consider the band assignment problem in dual band systems, where the base-station (BS) chooses one of the two available frequency bands (centimeter-wave and millimeter-wave bands) to communicate data to the mobile station (MS). While the millimeter-wave band offers higher data rate when it is available, there is a significant probability of outage during which the communication should be carried on the centimeter-wave band.  

In this work, we use a machine learning framework to provide an efficient and practical solution to the band assignment problem. In particular, the BS trains a Neural Network (NN) to predict the right band assignment decision using observed channel information. We study the performance of the NN in two environments: (i) A stochastic channel model with correlated bands, and (ii) microcellular outdoor channels obtained by simulations with a commercial ray-tracer. For the former case, for sake of comparison we also develop a threshold based band assignment that relies on the optimal mean square error estimator of the best band. In addition, we study the performance of the NN-based solution with different NN structures and different observed parameters (position, field strength, etc.). We compare the achieved performance to linear and logistic regression based solutions as well as the threshold based solution. Under practical constraints, the learning based band assignment shows competitive or superior performance in both environments.
\end{abstract}
\section{Introduction}
The large available bandwidth in the millimeter-wave (mmWave) frequency band can support the high data rates required for many applications. However, the hostile propagation conditions at high frequencies restrict the utilization of the mmWave band in cellular communication. Compared to the centimeter-wave (cmWave) band, signals in the mmWave band suffer from higher attenuation, higher diffraction loss, and are more susceptible to blockage, which reduces the reliability of the communication systems \cite{ProfMolischText,aliestimating}. To gain the advantages of both bands, next-generation wireless networks are anticipated to use both frequency bands \cite{andrews2014will,gonzalez2017millimeter}. This can enhance the signal coverage, system reliability and achievable data rates. 

The \emph{simultaneous} usage of the two bands might not be practical due to a number of limitations at the MS side, such as the limited processing capabilities, the constraint on transmission power, etc. Thus, the BS has to assign the MS to one of the two bands, in particular it has to switch the communication from the cmWave band to mmWave whenever the mmWave band is available or the other way around when the mmWave band suffers a blockage or other band propagation conditions. The band assignment (BA) problem is challenging, since the channel states in both bands might not be simultaneously observable at the BS side. In addition, using training signals over the two bands and frequent switching between the bands can be expensive and a waste of resources. To mitigate the aforementioned challenges, the BS can utilize the partial information, such as the channel state in one band, together with knowledge of BA "training data" to solve the BA problem.

Neural networks (NNs) have been successfully used in a number of wireless applications \cite{chen2017machine,o2017introduction}, as they proved to learn complex relations between the input data (features) and the output values (or labels). In this work we use NNs to provide a solution to the BA problem. This is motivated by the fact that the BA problem involves complicated relations between the observed features and the BA decision; these relations can be challenging to capture analytically. %In addition, using a NN we can utilize the prior knowledge about the channel states in the network and the past BA decisions. For these reasons the NN based solution have been explored for emerging applications.

 In the NN-based BA, the BS uses some of the features of the channel, such as the signal to noise ratio (SNR) in one band, the location of the MS, the signal delay, and/or the angle of departure to determine the band that could provide the highest \emph{data rate}. To train the NN, the BS can possibly use training data based on BA decisions at other locations and their associated observed features. Such information is relatively easy to collect as the MSs access the network regularly.
 \vspace{-2mm}
\subsection{Prior Work}
There have been a number of recent studies that considered the interplay between cmWave and mmWave bands. Refs.  \cite{Nitsche2015Steering,hashemi2017out,aliestimating} utilize the angular correlation in the two bands to provide an estimate of the Angle of Arrival at mmWaves, which can be used to reduce the beamforming complexity at mmWave. Furthermore, \cite{hashemi2017out} suggests using both frequency bands for data communication, and proposes a two-queue model to assign data to each band such that delay is minimized and throughput is maximized. Ref. \cite{ semiari2017joint} considers the downlink resource allocation in a network with a small cell BS, where the BS aims to assign the applications running on the MS's to the resources in the two bands. 
Our recent work \cite{burghal2018Rate} considers the band switching problem for an MS where the BS uses the observed channel states in the previous time-frames to predict the future one. Using a given channel model, the solution in \cite{burghal2018Rate} depends on the knowledge of the channel statistics and the path loss values, which might not be available in some practical situations. In addition, the channel states in previous time frames are usually not available during initial network access, or may convey relatively limited information for static or nomadic MSs.

Applying machine learning to solve wireless communication problems has gained considerable attention lately \cite{chen2017machine,o2017introduction}. For instance, recent studies use NNs to perform sequence detection \cite{farsad2018neural}, for jointly optimizing the encoding and the decoding in MIMO systems \cite{o2017deep}, and to provide power allocation in interference limited networks \cite{sun2017learning}. NNs have been also applied to learn the feature interrelation in wireless systems, Ref. \cite{Navabi2018Predicting} uses NNs to enable the BS to predict some unobserved features at the MS side, in particular, they try to infer the Angle of Arrival (AoA). 

\subsection{Contribution}
 Different from the above we consider the BA problem using the observed {\em instantaneous} features of the MS or its channel. Utilizing such information, the BS can reduce the required signaling, which improves the spectrum efficiency and reduces the latency in the system. The set of features available at the BS for the BA depends on the system setup, it \emph{may} include: the location of the MS, the received power (or data rate) in the cmWave band,\footnote{We use the SNR, signal strength and rate interchangeably when we refer to one of them as a feature, since we assume that we can use one of them to calculate the others, even though that might not be correct under some circumstances (e.g., interference-dominated channels).} the delay, and the Angle of Departure (AoD) of the main multi-path component (MPC) \cite{ProfMolischText}. 
 
 In this work, for ease of discussion and brevity, when we assess the impact of features availability, we focus on the features in cmWave band, i.e., switching from the cmWave band to the mmWave band, as the other cases are simple extensions. Furthermore, note that switching from the mmWave band to the cmWave band based on the observed SNR value is relatively safer than the opposite direction due to the relative reliability of the cmWave band \cite{gonzalez2017millimeter,burghal2018Rate}.
 
%  For instance, the BS may choose the appropriate band for an MS based on the SNR (or signal strength) of the messages that the latter uses to access the network in the cmWave band, this may speed up the initial access process, that otherwise may require unnecessary band switching and training.

 We use the NN framework, considering different NN structures and optimizing over different parameters. Additionally, we study the performance of the NN-based solution over different features combinations. To evaluate the performance of the NN-based solution we consider two environments, a stochastic and a ray-tracing based environments. The stochastic environment can provide initial assessment of the behavior of the proposed solution. As in \cite{burghal2018Rate}, we jointly generate the large scale fading in the two frequency bands. Furthermore, we utilize the mathematical tractability of the channel in this environment to develop a threshold-based BA (TBBA) scheme that we use to compare against the NN-based solution. The second environment is a data set generated by ray tracer to simulate the propagation conditions in the two frequency bands on a university campus. We use this environment to verify and extend our conclusions that we made based on the stochastic environment. In both environments, we use Linear Regression (LR) and Logistic Regression (GR) based solutions as benchmarks. Note that LR and GR are basic learning technique that capture the relation between the features and the BA. In summary, the contributions of this manuscript are threefold.
\begin{itemize}
\item We formulate the BA problem in a machine learning framework, and consider different NN configurations to optimize the performance of the proposed approach.
%\item We consider and design two relevant environments, a stochastic dual-band channel based on the band correlation, and  wireless channel in university campus that we simulated using a ray-tracer software. 
\item We study the impact of different features on the performance of the NN-based BA solution.% Which reveals interesting properties of such systems.
\item We compare the NN-based scheme to different benchmarks. In addition to the LR-based and GR-based BA, we develop the TBBA for meaningful comparisons in the stochastic environment.
\end{itemize}
The paper is organized as follows. Sec. \ref{sec:model} provides the system model. Sec. \ref{sec:NN} introduces the NN and highlights the set of features. Sec. \ref{Sec:LinPred} summaries the channel model, then provides the derivation of the TBBA rule, and discusses the performance of the schemes in the simulated stochastic channels and their generalization capabilities. Sec. \ref{sec:RT} includes the description of the ray-tracing environment and the performance of the learning-based schemes in that environment. Finally, Sec. \ref{Sec:Conc} provides concluding remarks.\allowdisplaybreaks
%%\vspace{-2 mm}
\section{System Model}\label{sec:model}
We consider a dual band cellular system, where the BS and the MS can operate in two frequency bands with center frequency $f_b$ and bandwidth $\omega_b$ in band $b \in \{c,m\}$, where $c$ and $m$ refer to the cmWave and the mmWave bands, respectively. Due to a number of practical limitations of the MS, we assume that data transmission occurs in a single frequency band at a time. The goal of the BS is to choose the band that results in the highest data rate. To focus on the basic problem, we consider a single user case, i.e., no scheduling or interference is considered; the multi-user case is left for future work. 

In this work, the BA procedure depends on the scheme. In NN-based schemes, the BS feeds the observed features, denoted by set $\mathcal{F}$, to the NN to produce the soft decision $\tilde{\mathcal{D}}$. Then it uses $\tilde{\mathcal{D}}$ to produce the BA decision $\mathcal{D}$. Note that since we have two distinct decisions, we can assume that $\tilde{\mathcal{D}} \in [0,1]$ and $\mathcal{D}\in\{0,1\}$, where $"1"$ refers to an assignment to the mmWave band. Thus we can view the problem as binary classification problem.  The BS uses a threshold $\gamma_{\rm L}\in [0,1]$ to map $\tilde{\mathcal{D}}$ to  ${\mathcal{D}}$, where we assume that ${\mathcal{D}} = 1$ when $\tilde{\mathcal{D}}>\gamma_{\rm L}$. The method we use to choose of $\gamma_{\rm L}$ is discussed in Sec. \ref{sec:NN}.

To train the NN, the BS uses a data set $\mathcal{A}^T = \{\mathcal{P}^T_1,...,\mathcal{P}^T_{N_T}\}$, where the superscript $T$ denotes training, and $N_T$ is the number of training examples. Each example point $\mathcal{P}^T_i$ is a features-label pair $(\mathcal{F}_i,\mathcal{L}_i)$, where $\mathcal{F}_i$ is the set of features of the $i$th example, and $\mathcal{L}_i\in \{0,1\}$ is the true label of that example, where $"1"$ refers to the case when the \emph{data rate} in the mmWave band is larger than the data rate in the cmWave band. We assume that $\mathcal{A}^T$ is available to the BS, for instance through previous decisions or an initial network training phase. However, the procedure to acquire $\mathcal{A}^T$ is out of the scope of the paper.

For clarity, we defer the discussion of the BA assignment procedure using the TBBA to subsection \ref{subsec:chAndThrshBA}.

\section{Neural Network and Features}\label{sec:NN}
\subsection{Neural Networks}
\begin{figure}
\centering
\vspace{-4 mm}
\includegraphics[width=17cm,height=5cm,keepaspectratio]{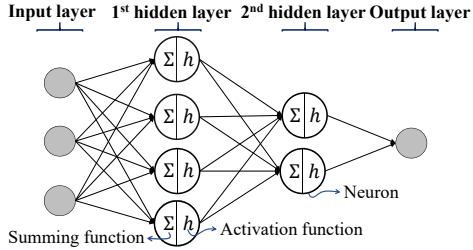}
\vspace{-9 mm}
\caption{Illustration of a neural network with two hidden layers.}
\label{fig:NNexample}
\vspace{-4 mm}
\end{figure}

Artificial neural networks have been successfully applied to many complex practical problems. An NN consists of one or more layers, each of which has a number of parallel neurons (nodes), see Fig. \ref{fig:NNexample}. The neuron performs a weighted combination of the input features and then passes it through a possibly non-linear transformation, also known as an activation function, e.g., a sigmoid function. The weights are determined during the training phase over a training set $\mathcal{A}^T$. where the goal is to minimize the prediction error of the label values at the output of the NN over the observed data points. 

 In this work we use feedforward NNs, where the output $\mathcal{D}$ depends solely on $\mathcal{F}$, i.e., the output at one instance does not influence the output at other time instances. To optimize the performance, we consider different NN structures. Note that while a large number of layers allows the NN to learn "difficult" aspects of the problem, this also increases the chances of over-fitting. In the simulation, we use several NNs structures with up to four hidden layers and up to $100$ nodes in total. To guard against over-fitting we use L2 norm regularization in the training phase with different regularization coefficients $\alpha$\cite{GooldFellow2016Deep}.%, which is sometimes referred to large variance, while for small networks we might suffer from large bias.%depending on the network structure and the size of the available data, the solution might suffer from bias or a large variance. %The number of layers and nodes per layer are hyper-parameters that need to be carefully chosen to prevent performance degradation due to these problems.

\subsection{Features}\label{subsec:Fet}
In this work we consider upto five features, i.e., channel characteristics that are used as input: ($\rm f_1$) the distance from the BS to the MS $d$ in meters, ($\rm f_2$) the angular position of the MS $\theta$ in rad, ($\rm f_3$) the received signal strength or the SNR in the cmWave band in dBm or dB (or equivalently the data rate in this paper), ($\rm f_4$) the delay of the main MPC in seconds, and ($\rm f_5$) the power of the main MPC in dBm, note that we refer to the MPC with highest power as the main MPC. 

Note that ($\rm f_1$) and ($\rm f_2$), i.e., $(d,\theta)$, represent the polar coordinates of the MS with respect to the BS, which may be estimated by localization techniques such as GPS. To extract ($\rm f_4$) large bandwidth might be required, for ($\rm f_5$) the use of antenna array is necessary. For $(\rm f_3)$, observation of the small-scale-averaged power are sufficient, because small-scale fading is usually averaged out by frequency diversity in large-bandwidth systems. This power can be observed at either the BS or the MS. As a result, the available features depend on the system implementations. In this work we consider several combinations of the above features.

As typically done in machine learning, we perform pre-processing to the features, such as normalizing the input features. We also use logarithmic scale for distances and power, as this may linearize their relation with one another.
\subsection{Training and Testing}
Let the set $\mathcal{A}$ denotes the entire data set we use in each environment, where each point $\mathcal{P}_i \in \mathcal{A}$ represents the features label pair $(\mathcal{F}_i,\mathcal{L}_i)$, where $i\in \{1,...,N\}$ and $N = |\mathcal{A}|$, $|.|$ denoting the cardinality operator. In the simulation we \emph{randomly} split $\mathcal{A}$ into a training set $\mathcal{A}^T$ and a testing set $\mathcal{A}^S$, where $\mathcal{A} = \mathcal{A}^T \cup \mathcal{A}^S$ and $\mathcal{A}^T \cap  \mathcal{A}^S = \emptyset$. We further split $\mathcal{A}^T$ into a training subset $\mathcal{A}_t^T$ and a validation subset $\mathcal{A}_v^T$, $\mathcal{A}_t^T \cap  \mathcal{A}_v^T = \emptyset$. More details about this are provided in the next sections.
During the training phase, as it is commonly used in binary classification problems, the performance of the NN is evaluated using Cross Entropy (CE) cost function, i.e., $$\bar{\mathcal{E}}_{{\rm CE},X} = \frac{1}{N_X}\sum_i^{N_X} (-\mathcal{L}_i log(\tilde{\mathcal{D}}_i)-(1-\mathcal{L}_i)log(1-\tilde{\mathcal{D}}_i)).$$ 
where the subscript $X\in\{T,V\}$ is used to identify, respectively, the training and the validation. We use Monte-Carlo cross validation to improve our estimate of the validation error, in which we repeat the random split of $\mathcal{A}^T$ to $\mathcal{A}_t^T$ and $\mathcal{A}_v^T$, and rerun the training and the validation. Then we choose the network structure and regularization coefficient that achieve the smallest \emph{average} $\bar{\mathcal{E}}_{\rm CE,V}$. In this paper, we focus on the average number of BA errors, i.e., we use 
\begin{align}\label{eq:error}
\bar{\mathcal{E}}_X = \frac{1}{N_X}\sum_i^{N_X} |\mathcal{D}_i-\mathcal{L}_i|.
\end{align} where the subscript $X\in\{S,V\}$ denotes, respectively, the testing and the validation. To choose $\gamma_{\rm L}$ we use the value in $[0,1]$ that results on the smallest \emph{average} $\bar{\mathcal{E}}_V$.\footnote{One may choose different performance metric, e.g., probability of error, and/or other methods to select $\gamma_{\rm L}$, such as the area under ROC curve \cite{GooldFellow2016Deep}, however we chose the above for clarity and ease of discussion.}

Finally note that we use similar training, cross validation and method to obtain the hard decisions for LR-based and GR-based techniques as well. %We use the error calculated over $\mathcal{A}^S$ for the chosen network structures to approximate the true performance of the scheme.
%we utilize the prior we consider several  explore the performance of the current problem, we use different architectures, such as cascaded network and combination of different combnation of activation function. 
\section{Stochastic Environment}\label{Sec:LinPred}
%To compare the performance of NN-based band assignment, 
Here we study the performance of the NN-based solution in stochastically generated channels. This will provide an initial assessment of the performance of the NN-based scheme with various feature combinations.%, which also help determining their impact on the BA solution.
\subsection{Channel Model and the TBBA Scheme}\label{subsec:chAndThrshBA}
Similar to \cite{burghal2018Rate}, we assume the channel model consists of the pathloss and the large-scale fading. For band $b \in \{c,m\}$, the SNR on a logarithmic scale (dB) is given by
\begin{align}
{\rm SNR}^{b} = P^{b}_{\rm tx} - PL^{b}- N^b_0 + S^{b},
\end{align}
where $P^{b}_{\rm tx}$ is the transmitted power, $PL^{b}$ is the path loss, $N^b_0$ is the noise level, and $S^{b}$ is a random process that represents the large-scale fading. 
 
With the assumption that the shadowing values in the two bands are jointly normal (on a logarithmic scale), the BS can use the observed channel state or rate in the cmWave band to infer their values in the mmWave band. Then, ideally it would assign the MS to the band with the largest \emph{rate}. Since the channel state in the mmWave band is not observable, we may assign the MS to the mmWave band, if the probability to achieve a larger rate in the mmWave band is greater than $\gamma_T\in [0,1]$, i.e.,
\begin{align}\label{eq:Arule}
\mathbb{P}(\mathbf{R}^m \geq \mathbf{R}^c | \mathbf{R}^c = r_c ) \geq \gamma_T,
\end{align}
where $\mathbf{R}^b$ is the rate in band $b$. Note that the left hand side of (\ref{eq:Arule}) is equivalent to the optimum mean square error estimator of the BA decision for given channel state in the cmWave band. Using the Shannon capacity equation, the rate in band $b$ is 
\begin{align}\label{eq:rate}
\mathbf{{R}}^b = \omega_b \log(1+\gamma_b'10^{\gamma'' S^b}).
\end{align} where $\gamma'' = 0.1$ and $ \gamma_b' = 10^{(P^{b}_{\rm tx} - PL^{b}- N^b_0)\times 0.1}.$ Using (\ref{eq:rate}) we can rewrite the probability in (\ref{eq:Arule}) as:
\begin{align}\label{eq:Cprob}
 & \mathbb{P}\bigg(\omega_m \log(1+\gamma_m' 10^{\gamma'' S^m})\geq r_c  \bigg| \omega_c \log(1+\gamma_c' 10^{\gamma'' S^c}) = r_c \bigg) \nonumber \\ & = \mathbb{P}( S^m \geq v_1 | S^c = v_0), \end{align}
 where $v_0 = \frac{1}{\gamma''} \log_{10}(\frac{1}{\gamma_c'}(\exp^{{r_c}/{\omega_c}}-1))$ and $v_1=\frac{1}{\gamma''} \log_{10}(\frac{1}{\gamma_m'}(\exp^{{r_c}/{\omega_m}}-1))$. With the assumption that $ S^m$ and $ S^c$ are jointly normal, it is enough to determine the conditional mean $\mu_{m|c}$ and variance $\sigma^2_{m|c}$ to calculate the probability in (\ref{eq:Cprob}), which can be shown to be 
 $$ \mu_{m|c} = \rho_{m,c} \frac{\sigma_m}{\sigma_c} v_0 ~~~{\rm and} ~~~ \sigma^2_{m|c} = (1-\rho_{m,c}^2)\sigma_m^2,$$
 where $\rho_{m,c}$ is the correlation coefficient of $S^m$ and $S^c$. Thus we have $$
 \mathbb{P}( S^m \geq v_1 | S^c = v_0) = \mathbb{Q}\bigg(\frac{v_1 - \mu_{m|c}}{\sigma_{m|c}}\bigg)\geq \gamma_T,$$
 where $\mathbb{Q}(.)$ is the Q-function \cite{abramowitz1964handbook}. Taking the inverse of Q-function, and rearranging the terms, we have 
 $$ v_1 \leq \mathbb{Q}^{-1}(\gamma_T)\sigma_{m|c}+\mu_{m|c} =  \mathbb{Q}^{-1}(\gamma_T)\sigma_{m|c}+\rho_{m,c} \frac{\sigma_m}{\sigma_c} v_0.$$ Solving for $v_0$, then the BS would assign the MS to the mmWave band if the following condition is satisfied:
\begin{align}\label{eq:ScThr}
S^c \geq \frac{\sigma_c}{\rho_{m,c} \sigma_m}\bigg(v_1 - \mathbb{Q}^{-1}(\gamma_T)\sigma_{m|c}\bigg),
\end{align} 
%Note that the right-hand side of the above condition is a function of $\gamma_T$ and the observed rate $r_c$ in the cmWave band: the former is a design parameter and the latter is function of the observed channel state in the cmWave band. Note that for the BS to make the decision, we assume that it knows the channel statistics and the path-loss values to the MS in both bands. Then the BS compares the observed shadowing value in the cmWave band as in (\ref{eq:ScThr}).
{which is a function of the observed rate $\gamma_T$, $r_c$, $v_1$ and the statistics of the environment. Here we set $\gamma_T = 0.5$ as it is the "natural" choice in this case, we provide further discussion about that in \cite{burghal2018Deep}. To calculate $v_1$, we need to calculate $\gamma_m'$, which in turn requires the knowledge of the pathloss.}

\subsection{The Data Set} \label{subsec:dataSetStoch}
 To generate the channel realizations in the two bands, we use the model suggested in \cite{burghal2018Rate}, with shadowing standard deviation $\sigma_b$, decorrelation distance $d^b_{\rm dcor}$ in band $b$, and a correlation coefficient $\rho_{c,m}$. We further assume that the pathloss follows a break point pathloss model\cite{ProfMolischText}, with a break distance $d_{\rm break}$ and a propagation exponent 2 for $ d \leq d_{\rm break}$ and $\epsilon$ for $d > d_{\rm break}$. Table \ref{Tab:SimEnvrionStc} summarizes the values used for generating the data set. 
%\begin{table}[!ht]
%\centering
% \begin{tabular}{|c|c|}
% \hline
% Variable & Value (cmWave/mmWave) \\
% \hline
% Carrier Frequency $f_b$  & 2.5/28 GHz \\
% \hline
%  Bandwidth $\omega_b$ & 10/100 MHz \\
% \hline
% Tx power & 15/15 dBm \\
% \hline
% $\epsilon$ & 4 \\
% \hline
% Break distance $d_{\rm break}$ & 30 m \\
% \hline
% $d_{\rm dcor}$ & 18/13\\
% \hline
% $\sigma_b$ & 5/7 dB\\
% \hline
%  $\rho_{m,c}$ & 0.75\\
% \hline
% Noise Spectral Density & -174 dBm/Hz\\
% \hline
% \end{tabular}
% \vspace{1mm}
%  \caption{Stochastic channel simulation configurations}
% \label{Tab:SimEnvrionStc}
%    \vspace{-7mm}
%\end{table}
\begin{table*}[!htb]
    %\caption{Global caption}
    \begin{minipage}{0.25\linewidth}
\centering
 \begin{tabular}{|c|c|}
 \hline
\bf Variable &\bf   Band c/m  \\
 \hline
 $\boldsymbol{f_b}$  & 2.5/28 GHz \\
 \hline
  \bf Bandwidth $\mathbf{\omega_b}$ & 10/100 MHz \\
 \hline
 $\boldsymbol{P^{b}_{\rm tx}}$ & 15/22 dBm \\
 \hline
 $\boldsymbol{\epsilon}$ & 4 \\
 \hline
  $\boldsymbol{d_{\rm break}}$ & 50 m \\
 \hline
 $\boldsymbol{d_{\rm dcor}}$ & 25/24\\
 \hline
 $\boldsymbol{\sigma_b}$ & 5/7 dB\\
 \hline
  $\boldsymbol{\rho_{m,c}}$ & 0.75\\
 \hline
 \bf Noise Spectral Density & -174 dBm/Hz\\
 \hline
 \end{tabular}
 \vspace{2mm}
  \caption{Stochastic channel simulation configurations}
 \label{Tab:SimEnvrionStc}
    \vspace{-9mm}
    \end{minipage}\hfill
    \begin{minipage}{.67\linewidth}
    \centering
 \begin{tabular}{|c|c|c|c|c|c|c|c|}
 \hline
\bf Feature / Combination & {\bf c-1 }   & \bf c-2   & \bf c-3   &\bf c-4    & \bf c-5    & \bf c-6  & \bf c-7    \\
\hline
 $\boldsymbol{d}$         &\checkmark    &\checkmark &           &\checkmark &\checkmark  &           & \\
 \hline
 $\boldsymbol{\theta}$    &\checkmark    &\checkmark &\checkmark &           &            &           & \checkmark \\
 \hline
\bf cmWave Power          &              &\checkmark &\checkmark &\checkmark &            &\checkmark & \\
 \hline \Xhline{2.4\arrayrulewidth}  
 \bf NN $\boldsymbol{\bar{\mathcal{E}}_S}$    &.227  & .18  &.183  &.193   &.265  &.194 & .399 \\
 \hline 
\bf GR $\boldsymbol{\bar{\mathcal{E}}_S}$     &.262  & .19  &.191 &.191   &.265  &.192 & .459 \\
 \hline
 \bf LR $\boldsymbol{\bar{\mathcal{E}}_S}$    &.262  & .194  &.192 &.195   &.264  &.193 & .459 \\
 \hline    \hline 
 \bf TBBA $\boldsymbol{\bar{\mathcal{E}}_S}$ &\multicolumn{7}{|c|}{.192}  \\
  \hline 
 \end{tabular}
 \vspace{3mm}
  \caption{Performance of the learning over the stochastic data under different feature availability. Note that on average $49.3$\% of the labels are $"1"$.}
   \vspace{-9mm}
 \label{Tab:ResultsStoch}
    \end{minipage} \hfill
\end{table*}

{ We assume that the BS is located at the center of a square cell with a side length of $500$ m, the data set consists of $2000$ data points, which correspond to $2000$ uniformly distributed MSs around the BS. In this data set, to simplify the simulation environment, we focus on three features: location of the MS $(d,\theta)$ and the power in the cmWave band. %The labels indicate whether the rates in the mmWave band are higher than the ones in the cmWave band for given MSs.

 In this environment, we use $65$\% of the data set for training. For the Monte-Carlo cross validation, we take around $20$\% of $\mathcal{A}^T$ for the validation subset $\mathcal{A}_v^T$. We generate $1000$ independent channel/cell realizations to assess the performance of the learning based BA and the TBBA in the stochastic environment.
\subsection{Performance} \label{subsec:perfStoch}
 We can view each of the $1000$ cell realizations as a different cell, which is reasonable in cellular system. Then for every realization we repeat the training, validation and then testing.  %Table-IV summarizes the performance. Although the numbers are different compared to the previous environment, we notice similar trend. Distance only, c-4, provides the worse performance, while the received power in cmWave, c-5, provides a relatively low error for NN and LR based approaches. Combination of distance and other information show slight improvement in general.

Table-\ref{Tab:ResultsStoch} summarizes the results for all seven feature combinations. The 5th, the 6th and the 7th rows show the test errors of the NN-based, the GR-based and the LR-based solutions, respectively. We emphasize that the displayed performance values by no means the optimal values, as we have considered a limited number of structures and parameters and performed a grid search over them. The last row in the table shows the performance of the TBBA. In the generated data set we have about $50.7\%$ of the labels are "0". As a result, assigning the cmWave band for all points would result in error equal $0.493$, we use this value as a reference for comparison. 

In general, we notice the learning techniques provide significant improvements over the cmWave-only BA, with an advantage to the NN-based scheme over the other schemes, as the NN is able to learn the non-linearity in the feature(s)/BA mapping. This can be observed in the performance for the first features combination (c-1), i.e., the location of the MS, where we notice that the NN-based BA outperforms the other learning approaches. Next, adding the received power in the cmWave band, (c-2), provides an evident performance gain for all learning approaches. In fact, it seems that any other combination with the power information would provide comparable performance, especially when we use the angle information as in (c-3). % We also notice that (c-3) shows a performance close to (c-2), which could indicate that in such stochastic environment, the angle has low impact on the performance, thus by eliminating it we can simplify the model. %and possibly improve the performance. 

To analyze the improvements in (c-2), (c-3) and (c-4) compared to (c-1), we study the performance using the distance-only feature in (c-5), power-only feature in (c-6) and angle-only feature in (c-7). While the bad performance in (c-7) is somewhat expected, as the received power is independent of angle on average, the power in the cmWave band seems to reveal more information about the BA than the distance. In fact, we notice that the performance in (c-4) is close to (c-6). This should not be surprising, as the shadowing is better captured with the received power in the cmWave band compared to the distance. In the light of this observation, we notice an improvement in (c-3) compared to (c-4), where the angle will provide an additional information to the power where we can identify cluster of similar BA decisions. In general, the performances of the NN-based scheme and both regression schemes are comparable in (c-5) and (c-6), indicating that a simple linear function could be used for BA in these two cases, which could be justified by the fact that in (c-5) the \emph{average} power could be approximated as a linear function of distance, while for (c-6) with the assumption of jointly normal channels, the linear function is an optimal estimator. 

{For the TBBA, Table \ref{Tab:ResultsStoch} shows that the TBBA provides an improvement compared to the cmWave-only BA. We notice that using combination (c-2), (c-3), (c-4) and (c-6), the learning based schemes can achieve similar performance to the TBBA, these have been achieved without providing the structure and the statistics of the channels. In fact, the NN is able to outperform the TBBA in (c-2) and (c-3), interestingly, with power only feature the learning based solutions are roughly as good as the TBBA.}
%>>>>>  We notice notice that the performance in c-%This simple result has a major impact in the way we need to design analyze the dual band systems, indicating the significance of the received signal strength in the cmWave to infer the channel state in 

%The table also shows the error values for the threshold base. We notice that with a fixed $\gamma_T$ at $0.5$ the scheme shows high error value, compared to the portion the "1" labels in the data set $0.238$. However, when we consider the distance for choosing $\gamma_T$, the error drops to $0.131$, which is better than the distance only for learning scheme but worse than   the performance of the NN for other combinations. 

 %We here use $\gamma_T = 0.5$, as it is the natural choice and it provides a reasonable performance according to our Monte-Carlo simulation grid search over several values of $\gamma_T$ (not shown here). 
\subsection{Generalization}
As an alternative to the method above, we might want to train the learning based approaches on a few cell realizations and apply it on other realizations, i.e., we are interested in transfer the learning experience without the need to go through the training phase again for each cell. In general, learning in stochastic environments is difficult, as the randomness increases the unpredictability of the labels and complicates the description of the mapping (if it exists) between the features and the labels. As a result, we expect a degradation in the performance as we try to generalize the learned relations.% as some of them are only applicable for training data. %when we train the learning based approaches on certain cell realizations and test it on others that are independent of the training cells. 

We use the same data set to study the generalizations of the solutions. However, we divided the $1000$ cell realizations into group of $50$ realizations. For each group, we use $30$ realizations for training, five for validation and $15$ for testing. We average the performance for the $20$ groups. We highlight here the results for three cases: for (c-1) we have $\boldsymbol{\bar{\mathcal{E}}_S}$ for NN, GR and LR respectively: $\{.27, .271, .271\}$, for (c-2) we have $\{.195,.195,.198\}$, and for (c-6) we have $\{.195, .196, .196\}$, and the TBBA has a test error value $.195$.\footnote{Note that the test error is different here compared to one that is shown in Table \ref{Tab:ResultsStoch}, as the test data sets are different in the two cases.} We notice a degradation in the performance in (c-1) and (c-2), as the learning approaches are not able to use some of the learned relations, that were deduced from the correlation between shadowing in the training cell realizations, for the BA over the test data set, as the latter  is independent of the training data set in this case. Interestingly, we notice that for (c-6), the learning based approaches provide similar performance as the TBBA, which indicates that they are good alternative to the TBBA for some feature combinations.
 
%However, it might be useful to draw conclsuion ab %for instance, the received power is not only a function pathloss to the received power and the selected band.

Finally, we conclude this section by emphasizing three limitations of the TBBA; first, it requires the knowledge of the exact pathloss to calculate $v_1$. Second, the correlation models and the shadowing distribution should be known to the BS. Third, as we noticed in \ref{subsec:perfStoch}, since the TBBA was derived based on an average metric, it does not exploit the available information in the given realization of the environment.}

%\vspace{-3mm}
\section{Simulated Campus Environment}\label{sec:RT}
\subsection{The Data Set}
\begin{figure}[!ht]
\centering
\vspace{-8mm}
\includegraphics[width=20cm,height=6.7cm,keepaspectratio]{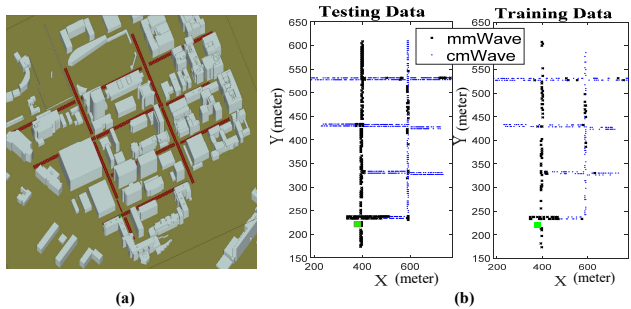}
\vspace{-17mm}
\caption{(a) Ray-tracing simulation environment. The green dot is the BS located above the rooftop, while simulated MSs are red routes. Gray objects represent the buildings. The light/dark green 3D polygons denote foliage features with different trees density. (b) Using 70\% of the data for testing, from left to right $\mathcal{A}^S$ and $\mathcal{A}^T$ with the corresponding labels.}
\label{Fig:Ray-tracing_USC}
 \vspace{-0mm}
\end{figure}

To assess the performance in a more realistic setting, we simulate the propagation channel in a campus environment by means of a commercial ray-tracing tool, Wireless InSite \cite{wiweb}. The input to the ray-tracer includes the 3D models of the buildings and the electromagnetic characteristics of the building materials as well as models of foliage. Wireless InSite performs {\em ray launching}, emitting rays (representing plane waves) from the transmitter into all directions, and following each ray as it interacts (reflection, diffraction, transmission) with the objects in the environment. The output is a list of parameter vectors that contains the power, propagation delay, the AoD and AoA, for each MPC. Simulation results have been compared to measurements in a variety of settings and shown to provide good agreement \cite{wiweb}. This simulation has been conducted based on the model of University Park Campus, University of Southern California (USC), which is shown in Fig. \ref{Fig:Ray-tracing_USC}-(a). The detailed simulation configurations are listed in Table \ref{Tab:SimEnvrionPara}. Simulation results in the same environment are used in \cite{AnsumanJSAC2014,LiTWC2018}.

The data set has about $1150$ points, i.e., $|\mathcal{A}| = 1150$, each point contains all the five features. The label that is associated with each point is whether the rate in the mmWave band is larger than the one in the cmWave band. To calculate the rate we use the Shannon capacity with bandwidth and noise spectral density that are shown in Table \ref{Tab:SimEnvrionPara}.

%\begin{table}[!ht]
% \vspace{-1mm}
%\centering
% \begin{tabular}{|c|c|}
% \hline
% Variable & Value (cmWave/mmWave) \\
% \hline
%$f_b$ & 2.5/28 GHz \\
% \hline
% Ant. Pattern & Isotropic \\
% \hline
% Ant. Polarization & Vertical \\
% \hline
% Tx power & 15/30 dBm \\
% \hline
% BS height & 45 m \\
% \hline
% MS height & 2 m \\
% \hline
% Maximal Diffraction & 2/1\\
% \hline
% Maximal Reflection & 10\\
% \hline
% \end{tabular}
 %\vspace{2mm}
 % \caption{Ray-tracing simulation configurations of USC campus}
  % \vspace{-6mm}
 %\label{Tab:SimEnvrionPara}
%\end{table}

%The set of observed features at the BS depends on the system setup and the content of the feedback, for instance with multi-antenna systems and/or large bandwidth, the BS could determine the AoD and the delay of the main multi-path component in the received signal, or it may implement localization techniques to estimate the location of the MS.\footnote{to attain the location of the MS, the BS may request the GPS reading from the MS.} In addition, the received power at cmWave could also be used as a feature. In this section we use several combinations of the above features as input to the NN. Then for each input $\mathcal{I}_i$ the NN makes the band assignment decision $\mathcal{D}_i$, which can be viewed as a binary decision, $\mathcal{D}_i \in \{1,0\}$, where one and zero, respectively, refer to the use mmWave and cmWave bands for communication.

Since acquiring a large number of data points may not be practical for the BS, using a large portion of the data set for training may produce misleading results. Here we use only 30\% of $\mathcal{A}$ for training. To apply the Monte-Carlo cross-validation method, we randomly choose 80\% of $\mathcal{A}^T$ for training and 20\% for validation. The network is then tested on $\mathcal{A}^S$, i.e., the remaining 70\% of the data set. Fig. \ref{Fig:Ray-tracing_USC}-(b) shows an example of the sets $\mathcal{A}^S$ and $\mathcal{A}^T$.

%\begin{figure}
%\centering
%\vspace{-2 mm}
%\includegraphics[width=9cm,height=5cm,keepaspectratio]{TrainTest}
%\vspace{-13 mm}
%\caption{Using 70\% of the data for testing, from left to right the test data set $\mathcal{A}^S$ and the training data set $\mathcal{A}^T$ with the corresponding labels, the BS is shown as square symbol.}
%\label{fig:TrainTest}
%\vspace{-7 mm}
%\end{figure}

\vspace{-0 mm}
\begin{table*}[!htb]
    %\caption{Global caption}
    \begin{minipage}{0.19\linewidth}
 \vspace{0mm}
\centering
 \begin{tabular}{|c|c|}
 \hline
 \bf Variable &\bf Band c/m \\
 \hline
$\mathbf{f_b}$ & 2.5/28 GHz \\
 \hline
 \bf Ant. Pattern & Isotropic \\
 \hline
\bf Ant. Polarization & Vertical \\
 \hline
 $\mathbf{P^{b}_{\rm tx}}$ & 15/30 dBm \\
 \hline
 \bf BS height & 45 m \\
 \hline
 \bf MS height & 2 m \\
 \hline
 \bf Max. Diffraction & 2/1\\
 \hline
 \bf Max. Reflection & 10\\
 \hline
 \end{tabular}
 \vspace{2mm}
 % \captionsetup{justification=centering}
  \caption{Ray-tracing simulation configurations of USC campus}
   \vspace{-9mm}
 \label{Tab:SimEnvrionPara}
    \end{minipage}\hfill
    \begin{minipage}{.755\linewidth}
\vspace{0mm}
\centering
%\begin{minipage}{1\linewidth}
 \begin{tabular}{|c|c|c|c|c|c|c|c|c|}
 \hline
\bf  Feature /Combination        & \bf c-1    & \bf c-2   &\bf  c-3   &\bf  c-4   & \bf c-5  & \bf c-6  &\bf  c-7 & \bf c-8  \\
\hline
$\boldsymbol{d}$       &\checkmark &\checkmark &\checkmark &           &           &\checkmark &           &           \\
 \hline
$\boldsymbol{\theta}$ &\checkmark &\checkmark &           &           &           &           &           &           \\
 \hline
  \bf cmWave Power   &           &\checkmark &\checkmark &\checkmark &\checkmark &           &\checkmark &           \\
 \hline
\bf Delay         &           &           &           &\checkmark &           &           &\checkmark &\checkmark \\
 \hline
 \bf AoD           &           &           &           &           &           &           &\checkmark &\checkmark\\
\hline \Xhline{2.3\arrayrulewidth} 
\bf Numb Layers/ $\boldsymbol{\alpha} $/ $\boldsymbol{\gamma_{\rm L}}$         & 2/.15/.45& 4/.15/.55& 1/.05/.5& 1/.1/.6& 2/.1/.35& 3/.5/.45& 4/.3/.6&3/.5/.55\\
 \hline
 \bf NN $\boldsymbol{\bar{\mathcal{E}}_{\rm S}}$  & .078& .061 & .072 & .074& .085& .182& .067& .093\\
 \hline 
  \bf  GR $\boldsymbol{\bar{\mathcal{E}}_{\rm S}}$  & .178& .062& .082& .081& .083& .183& .082& .182\\
 \hline 
 \bf   LR $\boldsymbol{\bar{\mathcal{E}}_{\rm S}}$  & .176& .078& .088& .078& .081& .178& .072& .188\\
 \hline 
 \end{tabular}
 \vspace{3mm}
  \caption{Performance of the learning techniques on ray-tracing data, under different feature availability, note that the percentage of points with labels equal to $"1"$ is approximately $30$\%.}
     \vspace{-9mm}
 \label{Tab:ResultsRay}
 \vspace{-0mm}
    \end{minipage} \hfill
\end{table*}
\subsection{Performance}
 We first point out that in this environment using the cmWave band only would result in an error equal to 0.3, i.e., the percentage of $"1"$ in $\mathcal{A}$ is 30\%. Table-\ref{Tab:ResultsRay} summarizes the results of the solutions. The last three rows show the performance of learning-based schemes when simulated over the described training set. The  the 7th row shows the number of layer (for an NN that consists of 100 nodes), the regularization coefficient $\alpha$ and the chosen threshold value $\gamma_{\rm L}$.\footnote{Again, we do not claim that the shown structures are the optimal choices.}% they are shown in the table for completion.} %For the reader's reference, in the 7th and the 8th row we show the validation error when we use the entire data set for training, i.e., $\mathcal{A} = \mathcal{A}^T$, with $20$\% of which used for validation, as it could be interesting to see the behavior of NN-based and LR-based solution for large data sets.

Combinations (c-1) and (c-8) show the cases when we use the location or the delay and AoD, these two are usually related as several localization techniques use the delay and AoD to determine the location. The performance in the two cases are comparable, even though we may not have Line of Sight (LOS) in all the cases. We also note that the NN-based approach significantly outperforms the other two approaches. 

Adding the power to the two combinations above, as in (c-2) and (c-7), improves the performance similar to the previous section, especially for both regression-based BA. In fact, in (c-7) their performance are slightly higher than the error for NN-based BA. 
The performance gain in (c-2) and (c-7) can be partially explained by the good results in (c-5) that uses the cmWave power only. For comparison, a scheme that only exploits the distance feature (c-6) shows relatively poor performance  for all the learning based schemes. This is consistent with our findings in the previous section, as the shadowing and the blockage have major impact on the quality of wireless channels especially in the mmWave band. Similar comparisons can be done with a delay-only (not shown in the table) scheme, which provides an improvement compared to distance only, with $\bar{\mathcal{E}}_{\rm S} = 0.17$ for NN-based, $\bar{\mathcal{E}}_{\rm S} = 0.165$ for GR-based and $\bar{\mathcal{E}}_{\rm S} = 0.166$ for the LR-based BA. This performance could be expected in that delay may reflect a more realistic "effective" distance, note that non-LOS links will show a longer delay even if they have similar geographic distance as their LOS counterpart. A combination of delay and distance with power, in (c-3) and (c-4), shows small improvement over power-only, however, they show significant improvement over distance-only and delay-only cases.  

Finally, we notice that in this environment,
the performance gaps between the NN based and other learning-based
BA solutions are in general larger than for the previous one.
Which suggests that in a more realistic environment, the NN is
especially useful.

%\begin{figure}
%\centering
%\vspace{-2 mm}
%\includegraphics[width=9cm,height=5cm,keepaspectratio]{OneRelNNPerf2}
%\vspace{-19 mm}
%\caption{One realization of training for c-4, the figure should also be compared with Fig.\ref{fig:TrainTest}, yet note that the plots are result of \emph{one} realization of $\mathcal{A}_t^T$.}
%\label{fig:TrainRel}
%\vspace{-5 mm}
%\end{figure}

%To give an illustrative example on the performance of the LR-based and the NN-based solution on the data, Fig. \ref{fig:TrainRel} shows the BA decisions on the entire data set after training over \emph{one} realization of the training subset $\mathcal{A}_t^T$. We notice that the NN is able to fit and predict labels of data points that might belong to a cluster of similar labels. This behavior needs to be used with care as it could also cause an over-fitting problem.

%-->As a concluding remark, we notice that in this environment, the performance gaps between the NN based and LR based BA solutions are larger than for the stochastic environment. This suggests that in a more realistic environment, with a more complex structure of shadowing and attenuation, NN is especially useful.
%\textcolor{blue}{To emphasize the practicality of the NN-based solution, we first notice that the NN structures in the 9th row are relatively small, which simplifies the training phase. In addition, note that acquiring some of these information is relatively easy, such as the SNR in the cmWave band. }
\vspace{-0mm}
\section{Conclusion}\label{Sec:Conc}
In dual-band systems, where the BS and the MS can communicate in either the cmWave or mmWave frequency bands, the BS should assign the MS to the appropriate band. In this paper we explore learning based approaches to provide a solution to the band assignment problem. We consider two environments to assess the performance of the proposed techniques and gain insight about the impact of different features: ray-tracing and stochastic environments. 
 
The performance of the schemes depends on the set of the available features. In both environments, the learning based approaches show impressive performance when the SNR in one band is known. In general, NNs (with relatively small number of nodes and hidden layers) show good performance using features that are relatively easy to acquire, such as the signal strength in one band and the delay of the main path. This indicates the practicality of the learning based schemes. The results also point out the importance of power correlation when studying and analyzing dual-band systems.%The NN based techniques have shown to adapt well to the conditions of the environment making them promising method many practical scenarios.
\vspace{-0mm}
\bibliographystyle{ieeetr}
\bibliography{reportBib}

\end{document}